  \documentclass[aps,prl,twocolumn,amsmath,amssymb,amsbsy,
superscriptaddress,floatfix,showpacs]{revtex4}

\usepackage[dvips]{graphicx}

\begin{document}

\title{Half--magnetization plateau stabilized by structural 
distortion in the antiferromagnetic Heisenberg model on a pyrochlore 
lattice}

\author{Karlo Penc}
\affiliation{
Research Institute  for  Theoretical Solid State  Physics   and
Optics, H-1525 Budapest, P.O.B.  49, Hungary}

\author{Nic Shannon}
\affiliation{
Department of Advanced Materials Science,
Graduate School of Frontier Sciences, University of Tokyo, 5--1--5, 
Kashiwahnoha, Kashiwa, Chiba 277--8851, Japan}
\affiliation{
CREST, Japan Science and Technology Agency, Kawaguchi 332-0012, Japan
}

\author{Hiroyuki Shiba}
\affiliation{
The Institute of Pure and Applied Physics, 6-9-6 Shinbashi,
Tokyo 105-0004}

\date{\today}

\begin{abstract}
Magnetization plateaus, visible as anomalies in magnetic 
susceptibility at low temperatures, are one of the hallmarks of 
frustrated magnetism.
We show how an extremely robust half--magnetization plateau can arise 
from coupling between spin and lattice degrees of freedom in a 
pyrochlore 
antiferromagnet, and develop a detailed symmetry of analysis of the 
simplest possible scenario for such a plateau state.
The application of this theory to the spinel oxides CdCr$_2$O$_4$
and HgCr$_2$O$_4$, where a robust half magnetization plateau has been 
observed, is discussed.
\end{abstract}

\pacs{
75.10.-b, 
75.10.Hk 
75.80.+q 
}

\maketitle

\begin{figure}[b]
  \centering
  \includegraphics[width=8.0truecm]{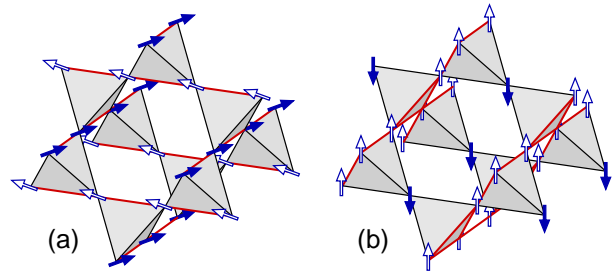}
  \caption{A section of the pyrochlore lattice, showing its 
  two--sublattice tetrahedral structure.  At  zero--magnetization, 
  tetragonal lattice distortion favors the spin configurations 
  of the type shown in (a), while at half--magnetization trigonal 
lattice 
  distortion favors the configuration (b).
  AF bonds are marked with black and 
  FM bonds with red lines. \label{fig:lattice}}
\end{figure}

Spinels, with chemical formula AB$_2$O$_4$, are ubiquitous among 
magnetic oxides.  Notable examples of such materials include 
Fe$_3$O$_4$, a system exhibiting frustrated charge order and 
ferrimagnetism \cite{anderson}, the $d$--electron heavy fermion 
compound
LiV$_2$O$_4$ \cite{urano}, and the frustrated $S=3/2$ antiferromagnet
ZnCr$_2$O$_4$ \cite{swc}.  
In all of these compounds, the B--site ion is magnetic, and
much of the beautiful strangeness seen in the behaviour of 
these compounds can be traced back to the fact that the 
B--site ions form an acutely frustrated {\it pyrochlore} lattice,
built entirely of corner--sharing tetrahedra (Fig.~\ref{fig:lattice}).

The geometric frustration of the pyrochlore lattice is so great 
that both the classical ($S=\infty$) and quantum ($S=1/2$)  
antiferromagnetic (AF)  Heisenberg models 
are believed to remain magnetically 
disordered down to $T=0$ \cite{moessner-chalker, canals-lacroix}.   
In real spinel oxides, however, the ground state 
degeneracy associated with the frustrated lattice geometry is usually 
lifted by a distortion of the lattice.   ZnCr$_2$O$_4$, for example, 
undergoes a transition from a paramagnet with cubic symmetry 
to a N\'eel ordered phase with tetragonal symmetry at $T=12K$.

Another, more progressive, means of reducing the ground state 
degeneracy of a frustrated AF is to apply a magnetic field.  Fields 
$h$ greatly in excess of the exchange coupling $J$ between spins will 
remove 
magnetic frustration altogether and cause the system to become 
ferromagnetic.  At intermediate fields $h \sim J$, 
frustrated AF's frequently undergo a succession of phase transitions,
with associated anomalies in their magnetic susceptibility.   
Where one particular state remains stable for a finite range of 
fields, a plateau is seen in the magnetization curve $M(h)$.   
Magnetization plateaus have been 
predicted to occur in both triangular lattice and Kagome AF's 
\cite{miyashita,andrey,momoi,mike}.  
In pure spin models, such plateaus occur
as an ``order from disorder effect'' where quantum or 
thermal fluctuations select one of many 
possible classical ground states\cite{mike2}.   
For this reason they are
usually very fragile, and relatively difficult to observe in 
experiment.

In this letter we consider the interplay between magnetic field,
spin, and lattice degrees of freedom in a Heisenberg antiferromagnet
on the pyrochlore lattice.
Our main result is that the coupling of applied magnetic field
to lattice distortion provides an extremely efficient mechanism 
for stabilizing a robust half--magnetization plateau, with exactly three up 
spins and one down spin in each tetrahedral sub--unit of the lattice.
In principle, many such states may arise; we develop a detailed
symmetry analysis of the simplest case, in which 
all of the tetrahedra which go to make up the lattice distort in 
the same manner.   

Our analysis is of direct relevance to 
spinel oxides such as ZnCr$_2$O$_4$, where the A--site ion is 
non--magnetic, and the octahedrally co--ordinated B--site 
ion has an exactly  half--filled $t_{2g}$ shell of $d$--electrons, 
giving rise to a spin $S=3/2$ moment.
And, indeed, just such a plateau has been observed in 
recent high field magnetization measurements on the 
closely related Cr spinels CdCr$_2$O$_4$ and HgCr$_2$O$_4$  
\cite{hiroyuki}.   

We take as a starting point the Hamiltonian
\begin{eqnarray}
 \mathcal{H} &=& \sum_{\langle i,j \rangle} 
  \left[
  J (1- \alpha_1 \rho_{i,j})
  {\bf S}_i {\bf S}_j
 + \frac{K}{2}  \rho_{i,j}^2 \right] 
- {\bf h} \sum_{i} {\bf S}_i \;,
\label{eq:H}
\end{eqnarray}
where the summation ${\langle i,j \rangle}$ runs over the nearest 
neighbor 
bonds of a pyrochlore lattice and $\rho_{i,j} $
is the change in distance between neighboring spins ${\bf S}_i$ and 
${\bf S}_j$, 
relative to the equilibrium lattice constant.
We assume the existence of a linear regime in which
exchange interactions and elastic energies 
depend only on the distance between lattice sites.
It is convenient to introduce a single dimensionless 
parameter $b = J \alpha^2/K$ to measure
the strength of the spin--lattice coupling.
As written, the antiferromagnetic exchange interaction $J$, elastic 
constant  $K$ and a spin--lattice coupling $\alpha$, are all taken to 
be positive.   

The pyrochlore is a bipartite network of corner sharing 
tetrahedra.
This means that, if we neglect coupling to the lattice, we can write 
the energy per spin as
\begin{equation}
 \mathcal{H} = 4 J\sum_{\mbox{tetr.}} 
  \left({\bf M} - \frac{{\bf h}}{8J}\right)^2 -\frac{h^2}{16J} + 
\mbox{const.} \;,
\end{equation}
where the sum runs over all the tetrahedra 
and \protect\mbox{${\bf M} = ({\bf S}_1 + {\bf S}_2 + {\bf S}_3 +  
{\bf S}_4)/4$}
is the magnetization per site.  Clearly, the energy can 
be minimized by ensuring ${\bf M} = {\bf h}/8$ in each tetrahedron.
However this local constraint does not select any one ground state, 
but rather a vast manifold of states.   If we consider classical 
spins, 
one angular variable remains undetermined per tetrahedron, 
and the magnetization is linear in $h$ right up the saturation 
field of $h = 8J$.

Coupling to the lattice provides a very efficient mechanism for
lifting this degeneracy.
Since bond energies vary  linearly with $\rho_{i,j} $, while elastic 
energies increase as $\rho_{i,j} ^2$, at any given value of magnetic 
field
the system can always gain energy by ordering the spins and 
distorting the lattice.
In this sense the Hamiltonian~(\ref{eq:H}) can be thought of as a
three dimensional generalization of the spin--Peierls problem.
The system gains the most energy from distorting bonds for
which ${\bf S}_i {\bf S}_j $ takes on its extremal values.  
For this reason, coupling to the lattice tends to favor collinear 
spin configurations and, for quantum spins, bond singlets (see e.g. \cite{becca}).
Our goal is to understand which states emerge from this
competition between applied field and frustrated AF interactions, 
and for what range of fields they are stable.

For the sake of simplicity, and in the spirit of earlier 
work~\cite{yamashita,oleg}, we shall restrict our analysis to
uniform spin and lattice order with crystal momentum $q=0$.
It is instructive to further simplify the problem by 
treating the spins as classical vectors, in which case
we can safely neglect all states which are odd under 
the inversion $\mathcal{I}_T$ which exchanges the two
tetrahedron sublattices.
Both of these approximations can be relaxed at will.

\begin{figure}[tb]
  \centering
  \includegraphics[width=6truecm]{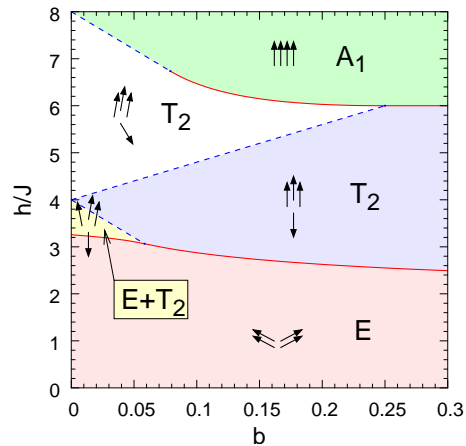}
  \caption{Phase diagram as a function of
  magnetic field $h$ and dimensionless coupling constant
  $b$.   Solid lines denote first and
  dashed lines second order transitions. Spin configurations and
  irrep of the order parameter in each phase is also shown. 
\label{fig:pd}}
\end{figure}

\begin{figure}[tb]
  \centering
  \includegraphics[width=7truecm]{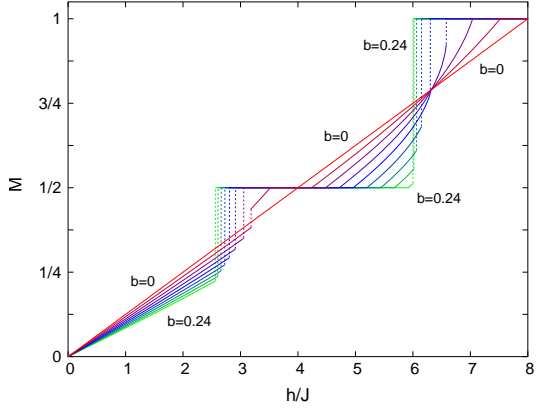}
  \caption{Magnetization as a function of magnetic field for $b=0$ 
(straight line) to $b=0.24$ (robust plateau) in steps of $0.03$.
\label{fig:hm}}
\end{figure}

Under these assumptions, the system must have four sublattice
order, and we can find the ground state of Eq.~(\ref{eq:H}) 
by a straightforward minimization of energy with respect to bond 
length.
This is equivalent to solving a Heisenberg model with 
biquadratic--bilinear 
terms\cite{kittel} :
\begin{equation}
 \mathcal{H} = \sum_{\langle i,j \rangle} 
 J \left[
   {\bf S}_i {\bf S}_j
   - b ({\bf S}_i {\bf S}_j)^2 \right] -  {\bf h} \sum_{i} {\bf S}_i
    \;.
   \label{eq:Hb}
\end{equation}
Our results are summarized in the phase diagram Fig.~\ref{fig:pd}, with
the corresponding magnetization curves shown in Fig.~\ref{fig:hm}. 
For small $h$, the lattice has overall tetragonal symmetry, with 
tetrahedra distorted so as to have two long ferromagnetic (FM) and four short canted 
AF bonds.
This is broadly compatible with the experimentally observed ground 
state of ZnCr$_2$O$_4$ \cite{swc}.  
In this regime the magnetization of the system remains linear,
but with reduced slope \protect\mbox{$M \approx h/(8J(1 + 2b))$}.

For $h \approx 3J$, and $b \gtrsim 0.05$ the system makes a 
first order transition into state with exactly 3 up and 1 down spins 
per tetrahedron, i.e.  $M \equiv S/2$, regardless of $h$.   
In this half--magnetization ``plateau'' phase, each tetrahedron has 
three long FM and three short AF bonds, giving rise to an overall trigonal 
lattice distortion.   For any finite value of $b$ the plateau is 
extremely broad.  Its width shrinks linearly
as $b \to 0$; for $h/J \leq 4 - 16b$ we find a transition into a 
coplanar 2:1:1 canted state with mixed ${\sf E}$ and ${\sf T}_2$ 
symmetry, and for $h/J = 4 + 8b$ a transition into a coplanar 3:1 canted phase 
with trigonal symmetry  for $h/J = 4 + 8b$.   Both transitions are 
of second order.  

Finally, for $8J >  h >  6J$, there is a transition into a fully 
saturated FM in which the lattice regains overall cubic symmetry.
In the absence of longer range exchange 
interactions, the two lowest lying spin wave branches of the FM phase 
are local in character and dispersionless.  
For $b < 3/38$, the transition from the 3:1 canted phase into the FM 
is of 
second order, and occurs on the line $h/J = 8 - 16b$.  For $b > 3/38$ 
the 
transition becomes first order, and for $b > 1/4$ it is energetically 
favorable to make a transition directly from the collinear plateau 
phase into the FM.

In order to understand {\it why} these particular phases are stable, 
we now turn to the symmetry analysis.   For classical spins, coupled
to uniform lattice distortion, we need only consider the symmetries $\mathcal{T}_d$ 
of a single tetrahedron \cite{footnote1}.
The bond variables $\rho_{i,j}$ , which describe changes in 
the length of the six edges of the tetrahedron, transform according to the ${\sf A_1}$, ${\sf E}$, and ${\sf T_2}$ irreducible representations (irreps) of $\mathcal{T}_d$:
\begin{equation}
 \left(
 \begin{array}{c}
  \rho_{{\sf A_1}}  \\
  \rho_{{\sf E},1}  \\
  \rho_{{\sf E},2}  \\
  \rho_{{\sf T_2},1}  \\
  \rho_{{\sf T_2},2}  \\
  \rho_{{\sf T_2},3}  \\
 \end{array}
\right)=
\left(
\begin{array}{cccccc}
\frac{1}{{\sqrt{6}}} & \frac{1}{{\sqrt{6}}} & \frac{1}{{\sqrt{6}}} &
 \frac{1}{{\sqrt{6}}} & \frac{1}{{\sqrt{6}}} & \frac{1}{{\sqrt{6}}} \\
\frac{1}{\sqrt{3}} & \frac{-1}{2\sqrt{3}} & \frac{-1}{2\sqrt{3}} &
 \frac{-1}{2\sqrt{3}} & \frac{-1}{2 \sqrt{3}} & \frac{1}{\sqrt{3}} \\
0 & \frac{1}{2} & -\frac{1}{2}  & -\frac{1}{2}  & \frac{1}{2} & 0 \\
0 & 0 & \frac{-1}{\sqrt{2}} & \frac{1}{\sqrt{2}} & 0 & 0 \\
0 & \frac{-1}{\sqrt{2}} & 0 & 0 & \frac{1}{\sqrt{2}}  & 0 \\
\frac{-1}{\sqrt{2}} & 0 & 0 & 0 & 0 & \frac{1}{\sqrt{2}} \\
\end{array}
\right)
\left(\begin{array}{c} 
 \rho_{1,2} \\
 \rho_{1,3} \\
 \rho_{1,4} \\ 
 \rho_{2,3} \\
 \rho_{2,4} \\
 \rho_{3,4} \\
\end{array}
\right)
\end{equation}
In the 
$\boldsymbol{\rho}_{\sf{E}}=\{ \rho_{\sf{E},1},\rho_{\sf{E},2} \}$ 
irrep (which includes tetragonal distortions of the lattice), 
opposing pairs of bonds deform with the same sense.
In the $\boldsymbol{\rho}_{\sf{T_2}}=\{ 
\rho_{\sf{T_2},1},\rho_{\sf{T_2},2},\rho_{\sf{T_2},3}
 \}$ irrep (which includes trigonal distortions 
), 
opposing pairs of bonds deform with the opposite sense.
Exactly analogous representations for spins can be obtained by 
substituting with ${\bf S}_i {\bf S}_j$ for $\rho_{i,j}$ and  
$\Lambda_{{\sf A_1}}$ for $\rho_{{\sf A_1}}$, etc.  

In terms of these variables, the Hamiltonian for a single tetrahedron 
embedded in the lattice is given by 
\begin{eqnarray}
  \mathcal{H} &=& 2 \sqrt{6} J \Lambda_{\sf{A}} 
      - 2 \alpha J 
   \left(
   \Lambda_{\sf{A}} \rho_{\sf{A}} 
   + \boldsymbol{\Lambda}_{\sf{E}} \boldsymbol{\rho}_{\sf{E}}
   + \boldsymbol{\Lambda}_{\sf{T_2}} \boldsymbol{\rho}_{\sf{T_2}}
   \right)
\nonumber\\
   && 
  +K \left(
   \rho_{\sf{A}}^2 + \rho_{\sf{E}}^2  + \rho_{\sf{T_2}}^2 \right)
   - 4 {\bf h}{\bf M} \;,
   \label{eq:Hagain}
\end{eqnarray}
where $\boldsymbol{\Lambda}_{\sf{R}} \boldsymbol{\rho}_{\sf{R}}$
and $\rho_{\sf{R}}^2=
\boldsymbol{\rho}_{\sf{R}}\boldsymbol{\rho}_{\sf{R}}$ are second 
order invariants 
of the $\sf{R}=\sf{A_1},\sf{E}$, and 
$\sf{T_2}$ irreps.
An analysis of Eq.~(\ref{eq:Hagain}) 
in the absence of magnetic field 
was given in Refs.~\cite{yamashita} and \cite{oleg}.
In all of the cases considered by these authors, only the 
${\sf E}$ irrep is relevant.
However, once a magnetic field is applied, both the 
${\sf A_1}$ and ${\sf T_2}$ irreps
have an important role to play.  
For classical spins, Eq.~(\ref{eq:Hagain}) reduces to
\begin{equation}
  E_0 = 2 J \bigl(\sqrt{6} \Lambda_{\sf{A_1}} 
   - b_{\sf{A_1}} \Lambda_{\sf{A_1}}^2
   - b_{\sf{E}} \Lambda_{{\sf E}}^2
   - b_{\sf{T_2}} \Lambda_{{\sf T_2}}^2 \bigr)
      - 4 {\bf h}{\bf M} \;.
   \label{eqn:E0}
\end{equation}
For pure nearest neighbour interaction [c.f. Eq.~(\ref{eq:H})]
\protect\mbox{$b_{\sf{A_1}}=b_{\sf{E}}=b_{\sf{T_2}} = b$}.   
In general, however, these parameters need not be equal.

\begin{figure}[tb]
  \centering
  \includegraphics[width=8truecm]{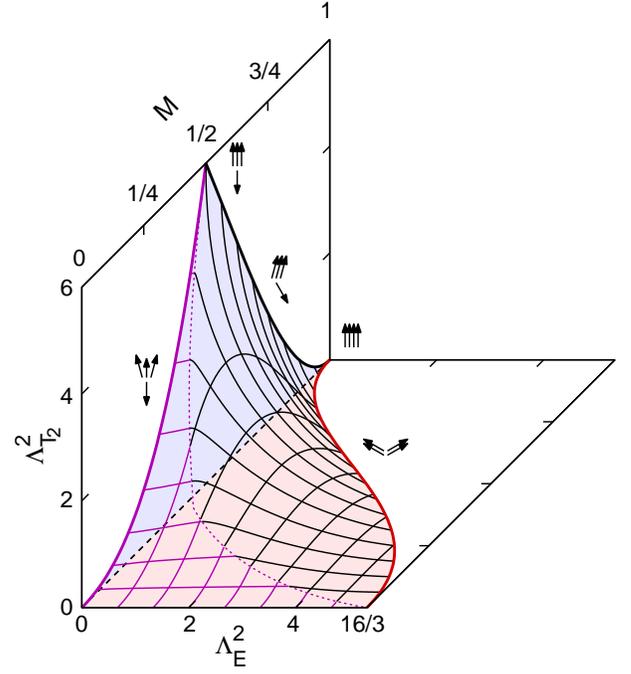}
  \caption{Maximal values of the second order invariants 
$\Lambda_{{\sf E}}^2$ and 
$\Lambda_{{\sf T_2}}^2$ for classical spins as a function of the 
magnetization per 
site $M$.  A ridge (dashed line) divides three dimensional from
coplanar spin configurations.  Spin ordering patterns are shown for 
the 
symmetrical cases of $\Lambda_{{\sf E}}^2=0$ and 
$\Lambda_{{\sf T_2}}^2=0$.   
\label{fig:me2t2}}
\end{figure}

Using the fact that $\Lambda_{\sf A_1}=8(M^2-1/4)/\sqrt{6}$,
we see that the lowest energy configuration at a given magnetization
is that for which 
$b_{\sf{E}} \Lambda_{{\sf E}}^2 + 
  b_{\sf{T}_2} \Lambda_{{\sf T}_2}^2$
takes on its maximal value. 
The surface of maximal values of these second 
order invariants is shown in Fig.~\ref{fig:me2t2}.
The limiting cases $\Lambda_{{\sf T_2}} \to 0$ or
$\Lambda_{{\sf E}} \to 0$ have a simple analytic form
\begin{eqnarray}
 \max_{\Lambda_{{\sf T_2}}^2 =0} \Lambda_{{\sf E_2}}^2  &=& 
 16(1-M^2)^2/3 \;,\\
   \max_{\Lambda_{{\sf E}}^2 = 0} \Lambda_{{\sf T_2}}^2 &=& \left\{
   \begin{array}{cc}
   \frac{32}{3} \left(1-M^2 \right)^2 \;,
    & \mbox{if $1/2 \leq M \leq 1$\;;}\\
   \frac{32}{3} M^2 (1+M)^2  \;,
    & \mbox{if $0 \leq M \leq 1/2$\;.}
   \end{array}
   \right.
 \end{eqnarray}
and the stability of the half--magnetization plateau originates in the sharp cusp 
in the maximal value of $\Lambda_{{\sf T}}^2$ as a function of $M$.
Provided that $b_{\sf{E}}<2b_{\sf{T_2}}$, this singular point
(which corresponds to the trigonal lattice distortion shown in 
Fig.~\ref{fig:lattice}b), minimizes the energy for a finite range of 
values of magnetic field, and the phase diagram is qualitatively 
that of Fig.~\ref{fig:pd}.   
Conversely,  for values of $b_{\sf{E}}>2b_{\sf{T_2}}$, ${\sf T_2}$ 
order is not realized for any value of $h$.
 
As the magnetization of the system increases, so will the average 
bond length, and for classical spins coupled to a uniform lattice 
distortion,
the volume of the unit cell is a monotonically increasing function of 
applied field 
\begin{equation}
  \frac{\delta V}{V} = \sqrt{\frac{3}{2}} \rho_{\sf A_1} 
  +\frac{1}{2}\rho_{\sf A_1}^2
  -\frac{7}{4} \rho_{{\sf E}}^2
  -\frac{3}{4} \rho_{{\sf T_2}}^2 \;.
\end{equation}
From this expression we learn that: (i) as 
$\rho_{\sf A_1} \propto \Lambda_{\sf A_1}$, the 
jumps in magnetization will have their counterpart in volume change;
(ii) the application of hydrostatic pressure 
will soften the ${\sf E}$ mode relative to the ${\sf T_2}$ mode, 
thus extending the region of the tetragonal phase.

The scenario which we have presented is the 
simplest under which lattice distortion can stabilize a magnetization
plateau in a spinel oxide.  
Needless to say, the situation in a real spinel
oxide may be much more complex than that described above.   
Naive estimates suggest that there are of order $1.3^{N/2}$ 
collinear states with  3 up and 1 down spins in each tetrahedron, 
where $N$ is the number of spins in the lattice.  
In principle, any of these may couple to phonon modes.
Arbitrarily complex exchange interactions, competing with 
arbitrarily complex elastic energies, may give rise to an arbitrarily 
complex plateau state --- or none at all.   
None the less, our model captures the essential features of the 
high field magnetization of CdCr$_2$O$_4$ and HgCr$_2$O$_4$ \cite{hiroyuki}.
For these Cr spinels $t_{2g}$ shell is full and the $e_g$ shell empty, 
so we may safely neglect the effects of orbital degeneracy \cite{motome}.
The theory may also be applicable to ZnCr$_2$O$_4$, although the high value of
$J$ in this compound makes verification difficult.

Chromium spinels exist as chalcogenides as well as oxides\cite{chalcogenides}.   
The chalcogenides have weaker and predominately FM interactions, and 
are therefore less likely  {\it a priori} to exhibit a half--magnetization plateau.
However the competition between FM and AF interactions is 
already present in oxide materials.   We have 
examined the role of next--nearest neighbour exchange $J_2$ within 
spin wave theory, the minimization of the energy of small clusters,
and an extension of the symmetry analysis presented above.
For AF $J_2$, $b_{\sf{T_2}}  > b_{\sf{E}}$, and the stability of 
the trigonal half--magnetization plateau phase is enhanced.   
FM $J_2$, on the other hand, drives the system towards 
lattice distortions (and associated plateau states) at finite $q$.
It is also worth noting that the strength and sign of exchange interactions 
in oxides and chalcogenides can be very sensitive to bond angle.
As written, the Eq.~(\ref{eq:H}) does not allow for 
the rotation of neighboring tetrahedra at fixed bond length.
Such modes will be important at finite $q$, and may lead to
a magnetostriction (i.e. decrease in the volume of the unit cell at in applied field).
The necessary extension of our theory to treat these cases is in 
principle possible, but lies beyond the scope of the present letter.   

We conclude with a few comments on the role of fluctuations.
In constructing a theory of a half--magnetization plateau stabilized 
by lattice distortion we have assumed static spin order.
Since both quantum and thermal fluctuations in frustrated 
magnets favour collinear spin configurations \cite{henley}, 
these will further contribute to the stability of the magnetization plateau.
If consider Eq.~(\ref{eq:Hb}) as an effective hamiltonian, the 
coefficient $b$ will have contributions fluctuation effects as
well as lattice distortion \cite{shiba}.   It may also have 
contributions of a purely electronic origin \cite{henning}.
However, we have performed monte carlo simulations of 
Eq.~(\ref{eq:H}) for classical spins in the absence of 
coupling to the lattice, and these suggest 
that any plateau stabilized by ``order from disorder'' effects alone will
be at least as fragile as that seen in the Kagome lattice \cite{mike}.
The remarkable width of the half--magnetization plateaus
observed in CdCr$_2$O$_4$ and HgCr$_2$O$_4$ --- 
which extend over many tesla --- leads us to the conclusion 
that lattice distortion plays a crucial role in these systems.

We are pleased to acknowledge helpful discussions with 
S.~W.~Cheong, P.~Fazekas, M.~Hagiwara, H.~Katori, M.~Matsuda, H.~Takagi, 
O.~Tchernyshyov, H.~Tsunetsugu, H.~Ueda, and K.~Ueda. 
We are particularly grateful to the authors of \cite{hiroyuki} for 
supplying us with their data prior to publication. 
We thank the support of the Hungarian OTKA T038162 and JSPS--HAS joint project.


\begin{thebibliography}{99}
\bibitem{footnote1} Formally, the point group of the pyrochlore lattice 
is $\mathcal{O}_h \equiv \mathcal{T}_d  \otimes \{1,\mathcal{I}_T\}$;
here by construction we work only with states even in $\mathcal{I}_T$.
\end{thebibliography}
\end{document}